\documentclass[]{emulateapj}

\newcommand{\msun}{M_{\odot}}
\newcommand{\mbh}{M_{\rm BH}}

\defcitealias{ina16}{IH16}

\shorttitle{Quenching of BH growth around the maximum mass}
\shortauthors{Ichikawa \& Inayoshi}

\usepackage{amsmath}
\usepackage{color}
\usepackage{url}
\usepackage{ulem}
\usepackage{multirow}
\usepackage{longtable}

\usepackage{graphicx}
\usepackage{epstopdf}

\def\msun{M_{\odot}}

\begin{document}

\title{Quenching of supermassive black hole growth around the apparent maximum mass\vspace{2.5mm}}

\author{Kohei~Ichikawa\altaffilmark{1,2,3} and
Kohei~Inayoshi\altaffilmark{1}
}
\affil{
$^1$ Department of Astronomy, Columbia University, 550 West 120th Street, New York, NY 10027, USA\\
$^2$ Department of Physics and Astronomy, University of Texas at San Antonio, One UTSA Circle, San Antonio, TX 78249, USA\\
$^3$ National Astronomical Observatory of Japan, 2-21-1 Osawa, Mitaka, Tokyo 181-8588, Japan\vspace{1.5mm}}

\email{k.ichikawa@astro.columbia.edu}

\begin{abstract}
Recent quasar surveys have revealed that supermassive black holes (SMBHs) 
rarely exceed a mass of $M_{\rm BH} \sim {\rm a~few}\times10^{10}~\msun$ during
the entire cosmic history.
It has been argued that quenching of the BH growth is caused by a transition of a nuclear
accretion disk into an advection dominated accretion flow, with which strong outflows
and/or jets are likely to be associated.
We investigate a relation between the maximum mass of SMBHs
 and the radio-loudness of quasars
with a well-defined sample of $\sim 10^5$ quasars at a redshift range of $0<z<2$, 
obtained from the Sloan Digital Sky Surveys DR7 catalog. 
We find that the number fraction of the radio-loud (RL)
quasars increases above a threshold of $M_{\rm BH}\simeq 2\times 10^{9}~\msun$,
independent of their redshifts.
Moreover, the number fraction of RL quasars with lower Eddington ratios 
(out of the whole RL quasars), indicating lower accretion rates, increases above the critical BH mass.
These observational trends can be natural consequences of the proposed scenario 
of suppressing BH growth around the apparent maximum mass of $\sim 10^{10}~\msun$.
The ongoing VLA Sky Survey in radio will allow us to estimate of the exact 
number fraction of RL quasars more precisely, which gives further insights to understand
quenching processes for BH growth.
\end{abstract}

\section{INTRODUCTION}

There is a fundamental question whether supermassive black holes (SMBHs) 
in the universe have a maximum mass.
The recent and on-going optical and near-infrared surveys 
of high-$z$ quasars have revealed that SMBHs 
with masses of $\mbh >10^9~\msun$ already exist at $z>6$
\citep{mor11, wu15, jia16, mat16}.
Since the $e$-folding time for BH growth in mass is only $\sim 40$~Myr,
SMBHs are likely to exceed $\mbh \sim 10^{11}$~$M_{\odot}$ significantly by $z \simeq 0$.
However, such SMBHs have not yet been observed in the local universe \citep{mcc11, kor13}.
More intriguingly, the surveys have also revealed that
there is a redshift-independent maximum mass limit at 
$M_{\rm max} \sim {\rm a~few}\times 10^{10}~\msun$ 
\citep{net03, mcl04a, ghi10, tra14}.

The origin of the maximum mass of SMBHs has been argued by several authors 
\citep[][hereafter \citetalias{ina16}]{nat09,kin16,ina16}
with a simple analytical model.
\citetalias{ina16} proposed that SMBHs exceeding 
$M_{\rm max}$ is prevented from growing by small-scale accretion physics, 
which is independent of the properties of their host galaxies or cosmology.
A high gas supplying rate from galactic scales into a nuclear region is required 
to form more massive SMBHs.
However, most of the gas is consumed by star formation 
in a gravitationally-unstable galactic disk
at large radii ($\gtrsim 100$~pc) well before reaching the nuclear region 
at $\lesssim 1$ pc \citep{tho05}, 
where the gas accretion rate results in $\lesssim 1\%$ of the Eddington accretion rate.
In such a low accretion rate, the gas flow at the vicinity of the SMBH never forms a 
standard geometrically-thin disk but transits into an advection-dominated accretion flow 
\citep[ADAF;][]{ichi77,nar94,nar95}.
Since the Bernoulli parameter of the ADAF is positive, 
strong outflows and jets are likely to be launched from the accreting system and 
shut off the BH feeding effectively \citep{IA99,bb99,sp01,HB02}.
Those outflows and jets interact with the ambient gas and form 
shock regions, from which radio emission can be produced by non-thermal electrons
(see a review by \citealt{yuan14}).
Thus, as a possible outcome of the quenching process, the largest SMBHs
would be associated with radio emission due to jets.

One key observational consequence of the above argument is that 
the number fraction of the radio-loud (RL) 
quasars (hereafter, RL fraction) in the high mass regime 
($\mbh >10^9$~$M_{\odot}$) would increase.
Observationally, RL quasars have systematically higher BH masses than 
their radio-quiet (RQ) counterparts \citep[e.g., ][]{lao00, mcl04, she08}, 
whereas a clear correlation between the radio-loudness parameter $R$ (defined later) 
and $M_{\rm BH}$ were not found \citep[e.g.,][]{ho02, woo02} 
up to $M_{\rm BH} \simeq 10^{10} \msun$.
\cite{bes05} first showed that the RL fraction increases with $\mbh$ using
the local ($z<0.5$) Sloan Digital Sky Survey (SDSS) DR2 type-2 active 
galactic nucleus sample for $\mbh \la 10^{9} \msun$.
This motivates us to extend the study to SMBHs with higher masses
 ($\mbh \la 10^{11}$~$M_{\odot}$),
where the critical BH mass theoretically expected is covered.

In this \textit{Letter}, in light of the large number samples of SDSS selected quasars,
we report the dependence of the RL fraction
as a function of $\mbh$ by expanding the BH mass range
with $10^{8} <\mbh<10^{11}$~$M_{\odot}$ and in the redshift range of $0< z < 2$.

\begin{table*}
  \begin{center}
    \small
    \caption{Number of (RL) quasars at each $M_{\rm BH}$ range}\label{tab:table1}
    \begin{tabular}{r|ccc|ccc|ccc}
      \hline
      \hline
      \multicolumn{1}{c}{} &
      \multicolumn{3}{c}{full sample} &
      \multicolumn{3}{c}{$0<z<1$ sample} &
      \multicolumn{3}{c}{$1<z<2$ sample} \\
      \hline
      $M_{\rm BH}$ range &
      $N_{\rm all}$ &
      $N_{\rm RL}^{(\rm min)}$ &
      $N_{\rm RL}^{(\rm max)}$ &
      $N_{\rm all}$ &
      $N_{\rm RL}^{(\rm min)}$ &
      $N_{\rm RL}^{(\rm max)}$ &
      $N_{\rm all}$ &
      $N_{\rm RL}^{(\rm min)}$ &
      $N_{\rm RL}^{(\rm max)}$
      \\
      (1) & (2) & (3) & (4) & (5) & (6) & (7) & (8) & (9) & (10) \\
      \hline
8--8.5 & 11330 & 400 & 661 & 8087 & 255 & 378 & 3241 & 145 & 283 \\
8.5--8.75 & 11651 & 445 & 607 & 5440 & 200 & 265 & 6209 & 245 & 342 \\
8.75--9 & 16028 & 691 & 796 & 4582 & 255 & 292 & 11444 & 436 & 504 \\
9--9.25 & 16283 & 781 & 838 & 2721 & 227 & 253 & 13562 & 554 & 585 \\
9.25--9.5 & 10767 & 680 & 699 & 1155 & 169 & 183 & 9612 & 511 & 516 \\
9.5--9.75 & 4440 & 376 & 381 & 347 & 62 & 67 & 4093 & 314 & 314 \\
9.75--10 & 1290 & 125 & 126 & 78 & 18 & 18 & 1212 & 107 & 108 \\
10--11 & 373 & 54 & 59 & 40 & 17 & 20 & 333 & 37 & 39 \\
      \hline
    \end{tabular}\\
  \end{center}
    Notes.--- Number of (RL) quasars at each $\mbh$ bin for the full sample (Column 2--4),
    the $0<z<1$ sample (Column 5--7), and the $1<z<2$ sample (Column 8--10).
    $N_{\rm all}$ represents the number of quasars.
    $N_{\rm RL}^{(\rm min)}$ and $N_{\rm RL}^{(\rm max)}$ are the number of RL quasars defined
    by the criterion (1) and (2) in \S2.2, respectively.
  \vspace{4mm}
  \end{table*}

\section{Sample}

\subsection{Quasar Catalog}

\begin{figure}
\begin{center}
\includegraphics[width=83mm]{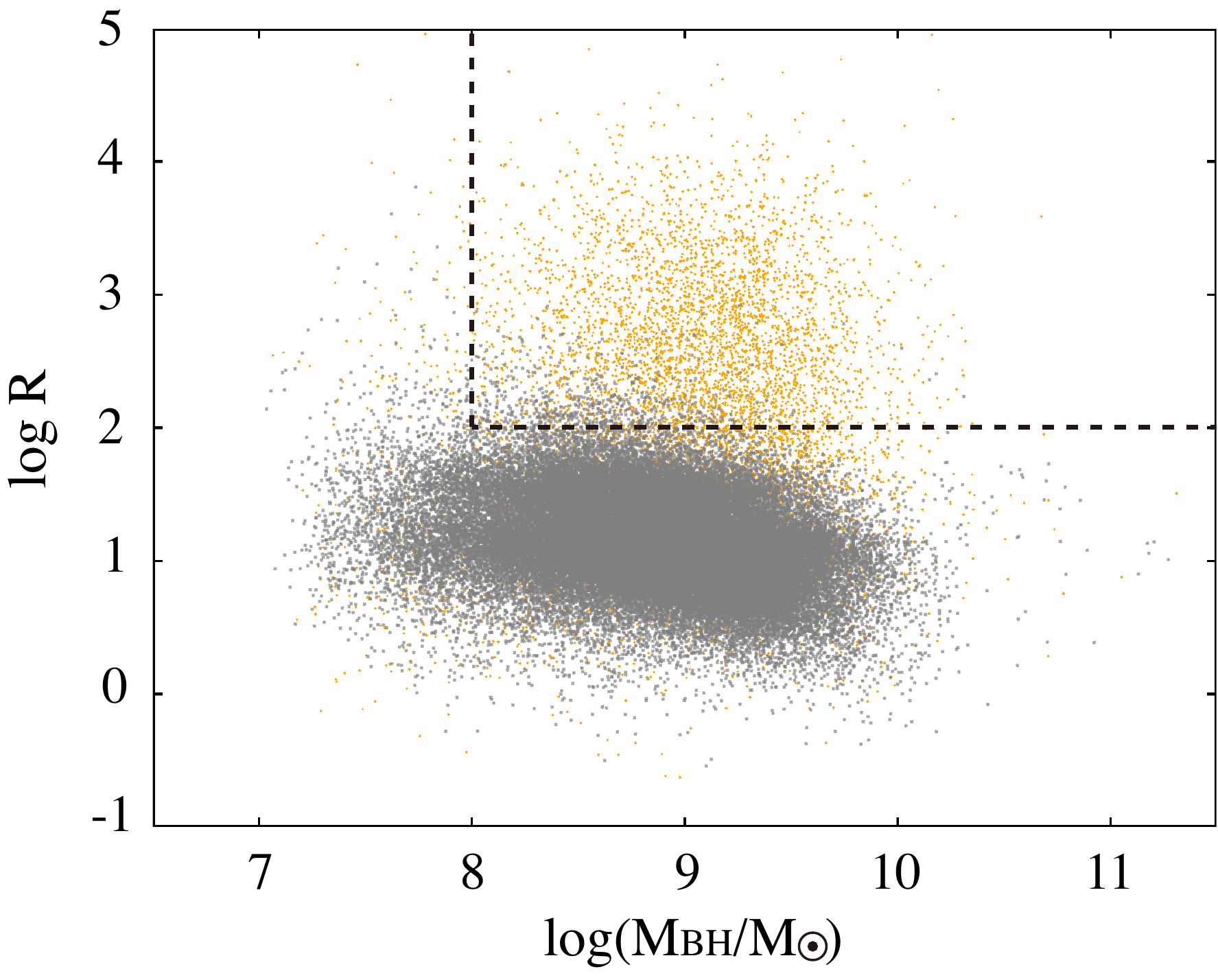}
\caption{
Distribution of radio-loudness parameters ($R$) for different BH masses.
Quasars with and without the radio detection
are shown by orange and gray dots, respectively. 
Dashed line gives the RL/RQ quasar boundary applied in this study; 
$R\geq 100$ and $\mbh \leq 10^8~\msun$ (see text in more detail).
}\label{fig:RvsMBH}
\end{center}
\vspace{4mm}
\end{figure}

The initial sample adopted in this letter is drawn from the SDSS DR7 Quasar Catalog 
\citep[][hereafter S11]{she11},
which contains 105783 quasars in the redshift range of $0.05 <z<5.5$.
The sources in the sample are brighter than the $K$--corrected,
$i$-band absolute magnitude $M_{i}(z=2)=-22$ which is normalized at $z=2$ \citep{ric06},
and have at least one broad emission line with full width at half-maximum larger
than 1000~km~s$^{-1}$. 
More detail sample selection and properties are described in S11.

In order to investigate the RL fraction of quasars as a function of $\mbh$,
we need data of the radio fluxes to measure the radio-loudness parameter $R$.
This value is defined by $R = f_{6{\rm cm}} / f_{2500}$,
where $f_{6 {\rm cm}}$ and $f_{2500}$ are the flux density at rest-frame 6~cm and 
2500~\AA, respectively \citep[e.g.,][]{kel89}, and conventionally adopted to separate 
samples into RL and RQ quasars at $R\sim 10$ \citep[e.g.,][]{kel89, kel94,ive02}.
Since the S11 catalog already contains the data of $f_{\rm 6cm}$ obtained from the FIRST survey 
\citep{bec95} and the positional information,
we first removed the sources whose FIRST radio flag$=-1$, 
indicating the source is not located in the FIRST footprint. This slightly reduces the sample into 99182.

In order to estimate BH masses, we here utilize H$\beta$ and MgII lines for quasars at $z<0.7$ \citep{ves06}
and $0.7<z<2.0$ \citep{she11}, respectively, as adopted in S11 as a fiducial choice of the BH estimates.
At $z>2.0$, CIV lines are commonly used to estimate BH masses.
However, we do not use the measurements with CIV lines because their mass 
estimation would have almost ubiquitous ambiguities due to 
non-virial wind component \citep{she08,tra12,mej16}. 
Finally, this selection reduces the number of sources to 75872 in the redshift interval of $0.05<z<2.0$.

 \begin{figure*}
\begin{center}
\includegraphics[width=88mm]{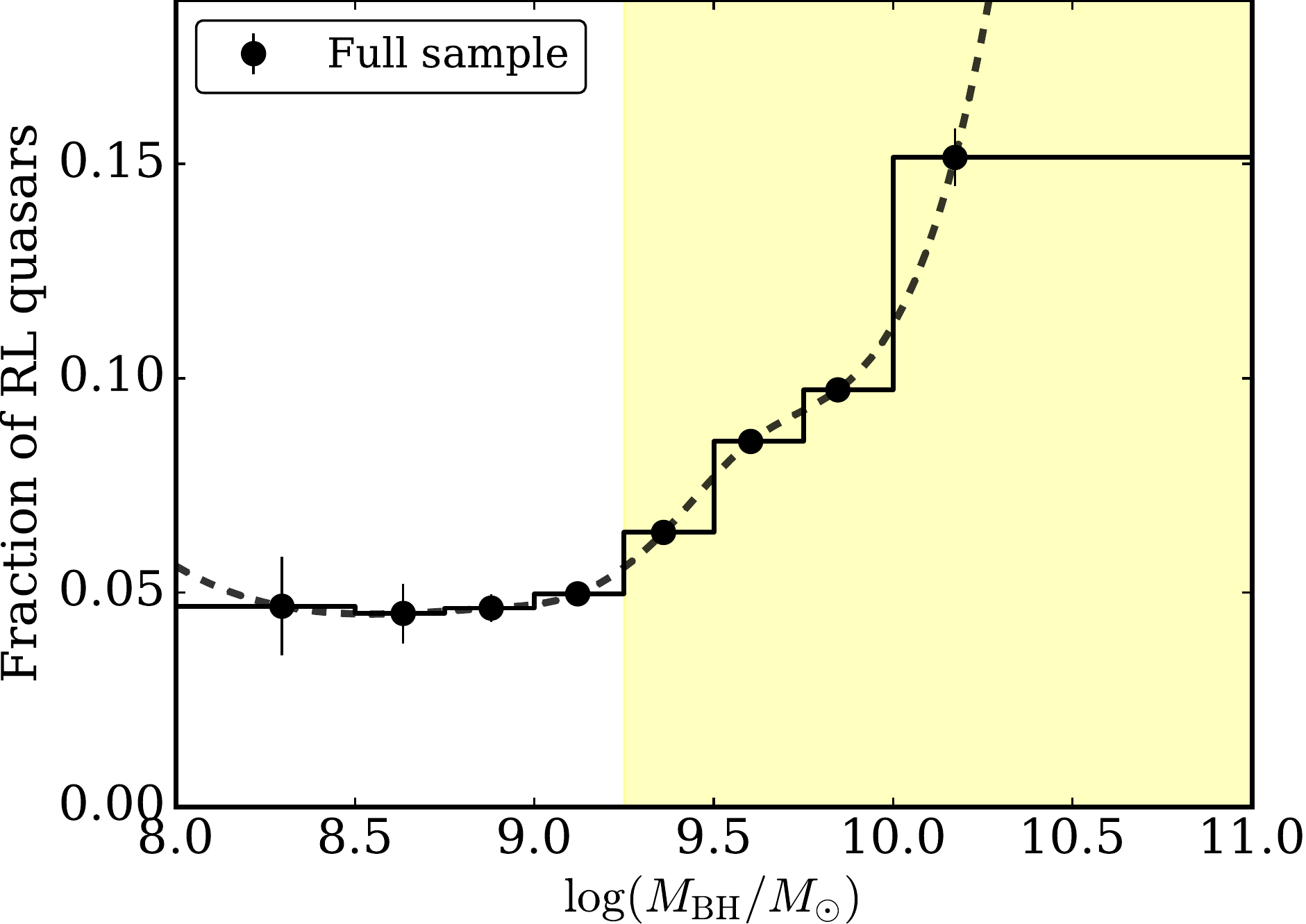}~
\includegraphics[width=86mm]{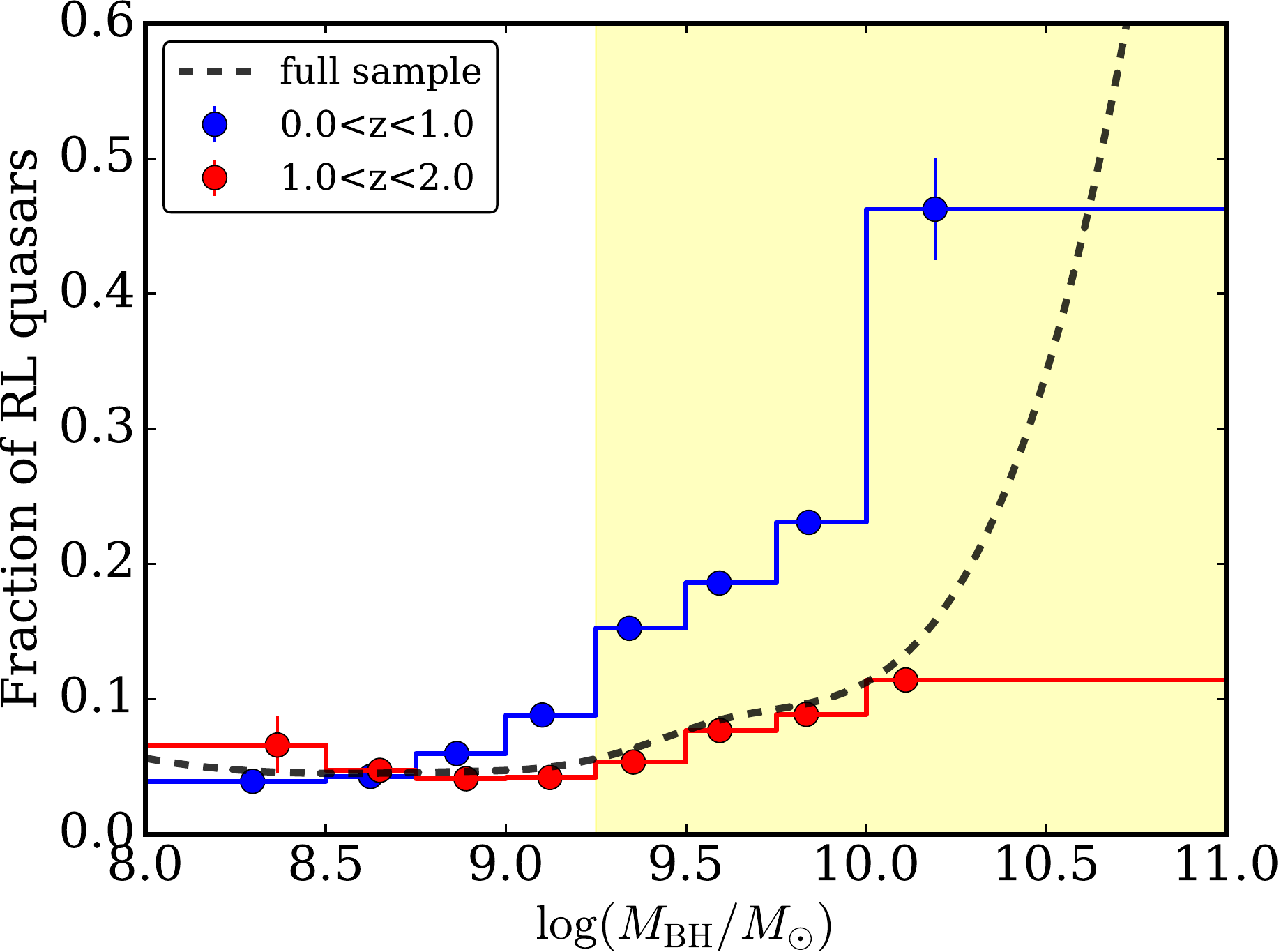}~
\caption{
RL fraction as a function of BH mass for the full sample at $0<z<2$ (black: left panel),
and for the sample at $z<1$ (blue: right panel) and $1<z<2$ sample (red: right panel), respectively.
The filled circle represents the average fraction of RL quasars at the median value of 
each BH mass bin in the histogram (solid).
We assign an upper limit of the RL fraction under the assumption that
all upper limit sources in the location above $R=100$ as RL quasars, and a
 lower limit assuming that all upper limit sources as RQ quasars.
 Dashed curve presents the cubic spline fitted line of the black filled points.
 The yellow area represents the region with $\mbh  > M_{\rm crit}=2\times 10^{9} \msun$.
}\label{fig:FRvsMBHall1}
\end{center}
\vspace{5mm}
\end{figure*}

\subsection{Radio loudness parameter and $M_{\rm BH}$}

Figure~\ref{fig:RvsMBH} shows the distribution of our quasar sample in the $\mbh$--$R$ plane.
Quasars with the detection and the non-detection in the FIRST band
are shown by orange and gray dots, respectively.
For the latter sample, the radio-loudness parameters are
estimated from the upper limits of the radio fluxes.
Because of the shallow sensitivity of the FIRST survey with a threshold of 1~mJy \citep{bec95},
the radio-loudness parameters $R$ have been measured for 7219 sources ($\simeq 9.5\%$) among all
the selected 75872 quasars.
As shown in Figure~\ref{fig:RvsMBH}, a significant fraction of the non-radio-detected quasars (gray dots)
distribute even above the conventional RL/RQ quasar boundary of $R=10$.
Therefore, we here adopt a RL/RQ boundary of $R=100$ instead of $R=10$, 
in order to reduce the contamination from non-radio-detected quasars and to give
a more conservative estimate of the RL fraction.
We also note that if we adopt $R>100$ as the RL/RQ boundary, only the sample size is reduced
without improving the contamination level.
In addition, we remove all the sources with $\mbh < 10^8~\msun$ because 
more than a half of the sources located at $\mbh < 10^8~\msun$ and $R>100$
are non-radio-detected sources,
and thus it provides a significant level of contamination
in those BH mass range even if we adopt the conservative RL/RQ boundary of $R=100$.
Finally, our sample contains 72162 sources with a mass range of 
$8.0< \log (M_{\rm BH}/M_{\odot})<11.0$ and a redshift range of $0.05<z<2.0$.
The number of quasars at each $M_{\rm BH}$ range ($N_{\rm all}$) is compiled at the
second column in Table~1.

We calculate the RL fraction based on different criteria for non-radio-detected quasars:
(1) those sources are always regarded as RQ quasars, and
(2) if the upper-limits of the radio-loudness parameters $R$ are higher than the RL/RQ boundary
of $R=100$, the sources are categorized as RL quasars.
The former (latter) cases give the lower (upper) limits of the RL fraction.
The lower and upper values of the RL fraction ($N_{\rm RL}^{(\rm min)}$) and 
($N_{\rm RL}^{(\rm max)}$) are shown in Table~1.

\section{Results}

The left panel of Figure~\ref{fig:FRvsMBHall1} shows the RL fraction as a function 
of $\mbh$ for the full sample ($0<z<2$). 
The RL fraction is almost a constant value of
$5~\%$ at $10^8\lesssim M_{\rm BH}/\msun \lesssim 2\times 10^{9}$.
This value is slightly smaller than the conventionally 
well-known value of 10\% \citep[e.g.,][]{kel89}
since we apply the conservative criterion of the RL/RQ boundary ($R=100$).
Above the critical mass of $\simeq 2\times 10^{9}~\msun (\equiv M_{\rm crit})$,
where the RL fraction becomes larger than their maximum values
at lower BH masses within the statistical errors,
the RL fraction begins to increase towards higher masses
and exceeds $15\%$ at $\mbh  >10^{10}~\msun$, 
which is three times higher than the constant fraction for lower masses.
The overall behavior of the RL fraction agrees with that expected from the theoretical model by \citetalias{ina16}.
Note that the enhancement of the RL fraction occurs slightly before the BH mass reaches $M_{\rm max}$.
We also confirm that the overall trend of the RL fraction above 
$M_{\rm crit}$ still holds as long as the RL/RQ boundary is set to $30<R<300$.

The right panel of Figure~\ref{fig:FRvsMBHall1} shows the RL fraction
for quasars at $0<z<1$ (blue), $1<z<2$ (red) and $0<z<2$ (black). 
For all the cases, the RL fraction is almost constant at $\mbh \lesssim M_{\rm crit}$ 
and begins to increase towards higher masses, as shown in Figure~\ref{fig:FRvsMBHall1}.
Therefore, the trend for the RL fraction does not depend on their redshifts, at least in the range of $0<z<2$.
We note that a slight enhancement of the RL fraction at $1<z<2$ seen 
in a mass bin of $10^8\leq \mbh /\msun \leq 10^{8.5}$ is due to 
the non-negligible level of contamination 
($138$ out of $283$ sources: see also Table~1) 
of the non-radio-detected sources.

\section{Discussions}

\subsection{Suggestion of ADAF Scenario}

\begin{figure}
\begin{center}
\includegraphics[width=85mm]{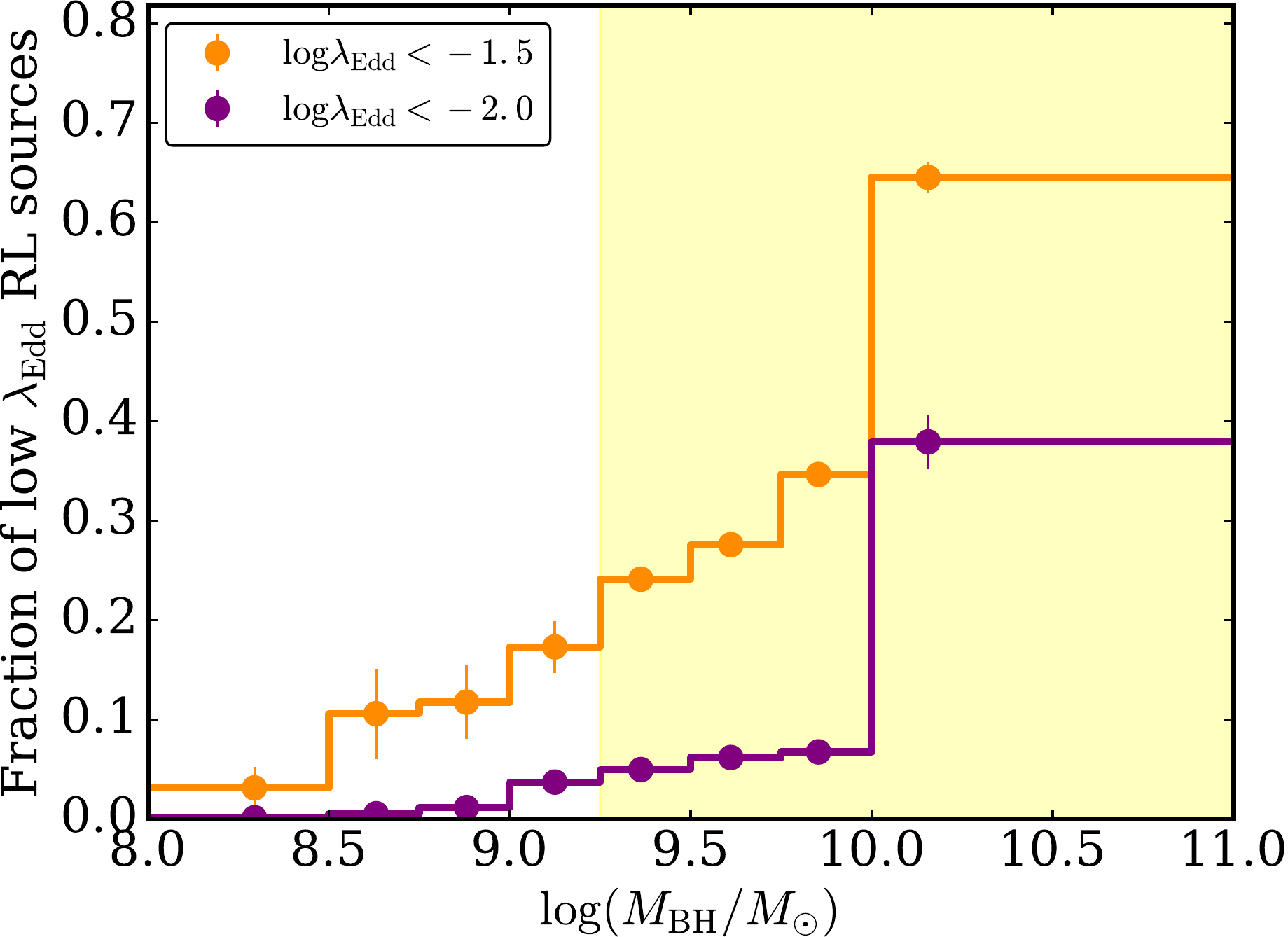}~
\caption{
Number fraction of RL quasars with lower values of the Eddington ratio 
$\lambda_{\rm Edd}$ as a function of the BH mass for the full RL quasar sample
for $\log \lambda_{\rm Edd}<-1.5$ (orange) and $\log \lambda_{\rm Edd}<-2.0$ (purple).
The filled circle represents the average fraction of RL quasars with lower $\lambda_{\rm Edd}$
at the median value of each BH mass bin in the histogram.
}\label{fig:EddvsMBHall}
\end{center}
\vspace{3mm}
\end{figure}

One scenario to explain the enhancement of the RL fraction above the critical mass
is due to a transition of an accretion disk to an ADAF (\citetalias{ina16}).
In order to test this theoretical model from another perspective, 
we investigate the Eddington ratios ($\lambda_{\rm Edd}=L_{\rm bol}/L_{\rm Edd}$)
of those RL quasars, where the bolometric luminosities $L_{\rm bol}$ are tabulated 
in the SDSS quasar survey \citep{she11} and $L_{\rm Edd}$ are the Eddington luminosities.
This is because the model also suggests that high-mass RL quasars should be fed with gas 
at low accretion rates, resulting in low Eddington ratios.
Figure~\ref{fig:EddvsMBHall} shows the number fraction of RL quasars
with $\log \lambda_{\rm Edd}<-1.5$ (orange) and $<-2.0$ (purple), respectively, 
{\it out of the whole RL quasars}.
The resultant number fraction for lower $\lambda_{\rm Edd}$ (i.e. lower accretion rates) 
also increases with BH masses, notably above $\mbh \sim M_{\rm crit}$, 
as expected from the theoretical model.
Our results would not prefer the possibility that a high RL faction is due to 
large-scale environmental effects of the host galaxies \citep{bes05}.
This is because the effects could explain the trend that the RL fraction increases 
with the BH masses, but do not describe the existence of the critical mass we found.

We note that several authors have discussed possible mechanisms 
leading to an upper limit on the BH masses. 
\cite{nat09} originally pointed out that the limit would be due to self-regulation 
of BH growth, related to the coevolution with the host galaxies.
Since gas clouds in halos with significantly higher masses than $\sim 10^{12}~\msun$ 
hardly cool, star formation in the halos would be reduced 
and thus the stellar masses are at most $M_\star \sim 10^{11}-10^{12}~\msun$ 
almost independently of redshifts ($0<z<5$) \citep{ro77}.
Assuming the mass ratios between SMBHs and their host galaxies of
$M_{\rm BH}/M_\star \sim 10^{-3}-10^{-2}$ in the local universe \citep[e.g.,][]{kor13},
the maximum BH mass would be estimated as $\sim 10^{10}~\msun$,
which is consistent with the apparent maximum mass of SMBHs.
However, since the BH/galaxy mass relation at $z\sim 0$ does not necessarily holds at high-redshifts
\citep{tra15,iho16,pac17},
the above mass limit may have a redshift-dependence 
(cf. the apparent maximum BH mass seems independent of redshifts).
In order to test the proposed scenario by \citetalias{ina16} of suppressing BH feeding
and distinguish from other scenarios, 
thus we need to investigate if the dependence of the RL fraction on the BH masses shown 
in Fig. \ref{fig:FRvsMBHall1} holds at higher redshifts.

As an alternative scenario, \cite{kin16} recently proposed that fragmentation of an AGN 
accretion disk due to gravitational instability would suppress the BH growth
at $\mbh \ga 5\times 10^{10}~\msun$.
This was because a gravitationally stable disk could not exist in the whole region
outside the inner-most stable circular orbit (ISCO).
However, since radiation pressure dominates near the ISCO and stabilizes the disk against 
its self-gravity, the critical mass due to fragmentation is boosted to $\mbh \ga 10^{13}~\msun$.
Since this mass is higher than the observed maximum mass of SMBHs by orders of magnitude 
(\citetalias{ina16}, \citealt{sha17}),
this suppression process of BH feeding could not explain the increase of the RL fraction around
$\mbh \sim 10^9-10^{10}~\msun$.

\subsection{Effects of the BH Spin on the RL Fraction}

BH spin potentially increases the radio luminosity from the AGN
because the rotational energy of the central SMBH is converted into the jet power \citep{bla77}. 
Previous observations have proposed that the most massive SMBHs 
have relatively higher positive spin \citep{tra14}. 
Assuming the Blandford-Znajek process, the jet luminosity is estimated as
$L_{\rm jet}\simeq \eta_{\rm jet}\dot{M}c^2$, where 
$\eta_{\rm jet} \sim 1.3 a^2$ \citep{tch15} and $a$ is the spin parameter.
On the other hand, the AGN bolometric luminosity is governed by
$L_{\rm bol}=\eta_{\rm rad} \dot{M}c^2$, where $\eta_{\rm rad}$ is the radiation efficiency.
The typical value is estimated as $\eta_{\rm rad}\simeq 0.1$ \citep{sol82}, which corresponds to 
mildly rotating BHs with $a \simeq 0.6$ and the highest value is $\eta_{\rm rad}\simeq 0.42$
for an extreme Kerr BH with $a =1$.
Therefore, the radio-loudness parameter can be estimated as 
$R\propto \eta_{\rm jet}/\eta_{\rm rad}\simeq 4.7~(3.1)$ for $a \simeq 0.6~(1.0)$.
Since the radio-loudness parameter $R$ hardly depends on the BH spin, 
the BH spin does not affect our conclusion.

We note that there is another caveat that the S11 quasar catalog 
used as the parent sample in this study would miss 
retrograde-spin ($a<0$) quasars in the high mass end
with $> 10^{10}$~$M_{\odot}$.
This is because those quasars produce relatively weaker UV excess, 
and therefore the color selections by SDSS could miss those weak 
UV excess quasars \citep{ber16}.
Therefore, we need to explore quasar samples with different color selections from that for SDSS quasars.
If we included those quasars in our sample, the RL fraction at the high mass end would increase 
because quasars with retrograde spins ($a\sim -1$) potentially provide higher radio-loudness parameters
($R\propto \eta_{\rm jet}/\eta_{\rm rad}\sim 30)$.

\begin{figure}
\begin{center}
\includegraphics[width=85mm]{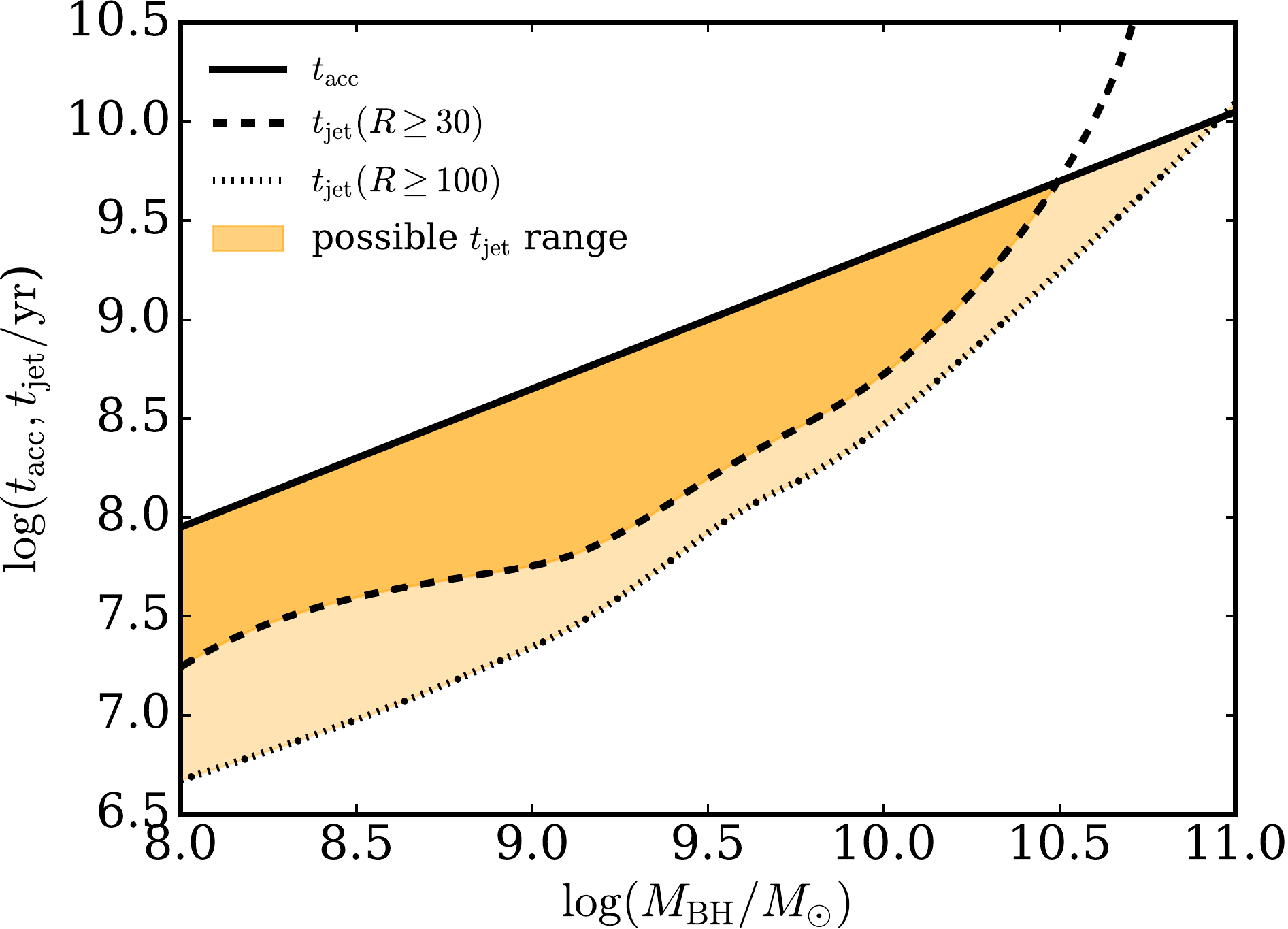}~
\caption{
Jet illuminating time ($t_{\rm jet}$) vs. mass-doubling time ($t_{\rm acc}$). 
The value of $t_{\rm acc}$ is derived from \citetalias{ina16} and 
$t_{\rm jet}$ is estimated using the spline fitting curve of the RL fraction shown in Figure~\ref{fig:FRvsMBHall1}.
The shaded area shows the possible range of $t_{\rm jet}$, which is in between
$t_{\rm acc}$ and $t_{\rm jet}(R\geq30~{\rm or}~100)$
estimated with two criteria for selecting RL quasars.
}\label{fig:tjetvsMBH}
\end{center}
\vspace{3mm}
\end{figure}

\subsection{Gradual Increase of the RL fraction above $M_{\rm crit}$}

One might argue that if the BH growth is ``turned off'' at certain critical mass,
the RL fraction increases drastically up to almost 100\%.
However, the RL fraction shown in Figures~\ref{fig:FRvsMBHall1} increases \textit{gradually}
 above the critical mass of $M_{\rm crit} \simeq 2\times 10^{9}~\msun$.
This fact suggests that the quenching process of BH growth would not occur instantaneously, 
but take certain amount of time and occur episodically until the disk state transits
to an ADAF state with the jet emission.
The observed RL fraction for a BH mass bin can be interpreted as the ratio of the
jet illuminating time $t_{\rm jet}$ (i.e., the disk is in an ADAF) to the time when
BHs exist in the mass bin.
Thus, the RL fraction is written as  $f_{\rm RL} \simeq t_{\rm jet}/ (t_{\rm acc} + t_{\rm jet})$,
where $t_{\rm acc}$ is the mass-doubling time (i.e., the disk in a standard disk). 
Figure~\ref{fig:tjetvsMBH} shows the possible range of
 $t_{\rm jet} \simeq t_{\rm acc}f_{\rm RL}/(1 - f_{\rm RL})$
as a function of $\mbh$, where the mass-doubling time is estimated as
$t_{\rm acc} \simeq 4.5\times 10^8 (\mbh/10^9~\msun)^{0.7}$~yr
\citep{tho05} and $f_{\rm RL}$ is estimated from the
dashed curve in Figure~\ref{fig:FRvsMBHall1}.
Since our study uses the criterion of $R\geq 100$ for selecting RL quasars, the derived value of
 $t_{\rm jet}(R\geq100)$ gives the lower limit of $t_{\rm jet}$ as shown in dotted curve. 
 We also over-plot the value of $t_{\rm jet}$ with $R\geq 30$ for choosing RL quasars,
 which gives the relatively larger value than that of $t_{\rm jet}(R\geq 100)$.
Figure~\ref{fig:tjetvsMBH} also shows that 
$t_{\rm jet}(R\geq 30)$ becomes larger than $t_{\rm acc}$ at $\mbh \gtrsim 10^{10.5}~\msun$.
That is, more than $50\%$ of nuclear disks are in ADAF states, 
in which the SMBHs would be prevented from increasing its own mass significantly.

\subsection{Exact Value of the RL Fraction}

The exact value of the RL fraction at each BH mass
is still unknown because of the shallower radio band sensitivity 
(see also Figure~\ref{fig:RvsMBH}).
This will be improved after the completion of the ongoing VLA Sky Survey 
(VLASS; \url{https://science.nrao.edu/science/surveys/vlass/vlass}), 
which covers the SDSS survey area
with the sensitivity down to 0.12 mJy at S-band (2--4~GHz),
giving us the almost one order of magnitude deeper photometry
compared to the current FIRST sensitivity of 1~mJy at 1.4~GHz.
While the frequency coverage is slightly different, this survey allows us to
discuss this argument with the conventional RL/RQ boundary of $R=10$,
instead of our conservative criterion of $R=100$.

\acknowledgments
We thank Zolt\'an Haiman, Benny Trakhtenbrot, and Kazumi Kashiyama for useful discussions.
This study is partially supported by the Grant-in-Aid for JSPS fellow for young researchers (KI) 
and by the Simons Foundation through the Simons Society of Fellows (KI).

\end{document}